\documentclass{article}
\usepackage[utf8]{inputenc}
\usepackage{amsmath}
\usepackage{nicefrac}
\usepackage[affil-it]{authblk}
\usepackage{booktabs}
\usepackage{changepage}
\usepackage{multirow}
\usepackage{tikz}

\usetikzlibrary{decorations.pathreplacing,decorations.markings,patterns}

\tikzset{
  on each segment/.style={
    decorate,
    decoration={
      show path construction,
      moveto code={},
      lineto code={
        \path [#1]
        (\tikzinputsegmentfirst) -- (\tikzinputsegmentlast);
      },
      curveto code={
        \path [#1] (\tikzinputsegmentfirst)
        .. controls
        (\tikzinputsegmentsupporta) and (\tikzinputsegmentsupportb)
        ..
        (\tikzinputsegmentlast);
      },
      closepath code={
        \path [#1]
        (\tikzinputsegmentfirst) -- (\tikzinputsegmentlast);
      },
    },
  },
  mid arrow/.style={postaction={decorate,decoration={
        markings,
        mark=at position .5 with {\arrow[#1]{stealth}}
      }}},
}

\title{Maximum likelihood estimation for disk image parameters}
\author{Matwey V. Kornilov\thanks{matwey@sai.msu.ru}}
\affil{\small{Sternberg Astronomical Institute, Lomonosov Moscow State University, \\ Universitetsky~pr. 13, Moscow 119234, Russia}}
\affil{\small{National Research University Higher School of Economics, \\ \nicefrac{21}{4}~Staraya Basmannaya Ulitsa, Moscow 105066, Russia}}
\date{}

\begin{document}

\maketitle

\begin{abstract}
We present a novel technique for estimating disk parameters~(the centre and the radius) from its 2D image.
It is based on the maximal likelihood approach utilising both edge pixels coordinates and the image intensity gradients.
We emphasise the following advantages of our likelihood model.
It has closed-form formulae for parameter estimating, requiring less computational resources than iterative algorithms therefore.
The likelihood model naturally distinguishes the outer and inner annulus edges.

The proposed technique was evaluated on both synthetic and real data.
\end{abstract}

\section{Introduction}
Circles and disks are among the most basic geometric primitives.
Therefore, the detection and fitting problems are widespread over different fields:
microwave engineering~\cite{Kasa1976},
particle physics~\cite{Karimaki1991,Crawford1983},
pattern recognition~\cite{Lin2004},
quality control~\cite{Landau1987},
robotic systems~\cite{Coach2003,Nunez2008,Zhang2006},
and others.
Since a disk edge is a circle, detecting a circle and detecting a disk are closely related problems.
In this paper, we apply disk fitting in the experimental astronomy field.

The techniques may be coarsely divided into the two following classes.
First, convolutional image-based techniques such as
the circle Hough transform~\cite{Ballard1981,Illingworth1987},
or the phase-coded annulus~\cite{Atherton1999}.
They work with a 2D image represented as an array of pixels.
Second, point-based statistical techniques, for instance, maximal likelihood estimation.
Here, input data is a list of points presumable residing on a circle.
A review of different approaches is given, for instance, in~\cite{Zelniker2006}.

In this work, we limit to fit only a single one disk or annulus per an image frame,
so that the 2D image array can be straightforwardly transformed to the list of points by a gradient filtering followed by thresholding such as Otsu thresholding~\cite{Otsu1979}.
As soon as the maximal likelihood fitting follows the gradient filtering, we also use normalised gradient vectors during the likelihood fitting.

The paper is organised as the following.
A gradient-based maximal likelihood estimation with an analytic solution is derived in Section~\ref{sec:mle}.
In Section~\ref{sec:eval} we describe full detection pipeline and evaluate it on synthetic data.
In Section~\ref{sec:pupil} we show how the expectation minimisation technique applied on top of maximal likelihood can recognise an astronomical telescope entrance pupil.
The major results are discussed in the Conclusion.

\section{Gradient-based likelihood model}
\label{sec:mle}

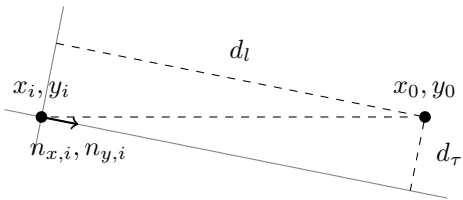
\begin{figure}
\centering
\begin{tikzpicture}
\begin{scope}[rotate=-11.31]
\draw [thin,gray] (-0.5,0) -- (5.5,0);
\draw [thin,gray] (0,-0.5) -- (0,1.5);
\node (i) at (0,0) [label=above:{$x_i, y_i$}] {};
\node (o) at (5,1) [label=above:{$x_0, y_0$}] {};
\filldraw [] (0,0) circle (2pt) (5,1) circle (2pt);
\draw [dashed] (0,1) -- node [label=above:{$d_{l}$}] {} (5,1) -- node [label=right:{$d_{\tau}$}] {} (5,0);
\draw [dashed] (i) to (o);
\draw [thick, ->] (0,0) -- (0.5,0) node [label=below:{$n_{x,i}, n_{y,i}$}] {};
\end{scope}
\end{tikzpicture}
\caption{
\label{fig:1}
An explanation for deriving equation~(\ref{eq:3}).
Here, $d_{l}$ is the signed distance between the point $x_0,y_0$ and the line crossing through $x_i,y_i$ in the direction orthogonal to the unit vector $n_{x,i},n_{y,i}$; $d_{\tau}$ is the signed distance for the line collinear to the unit vector.
}
\end{figure}

When a table of noisy $(x,y)$-pairs is known,
under a specific natural assumption we may derive the following log-likelihood loss function to estimate the centre and the radius for a circle:
\begin{equation}
\label{eq:1}
\ln p(x_i, y_i | \theta) = -\frac{1}{2}\ln 2 \pi \sigma - \frac{1}{2\sigma^2} \sum^{N}_{i=1} \left(\sqrt{(x_i - x_0)^2 + (y_i - y_0)^2} - R\right)^2,
\end{equation}
where $\theta$ is the set of distribution parameters: the centre $x_0,y_0$, the radius $R$.
The measured points are denoted by $x_i,y_i$, $N$ is the total points number.
For instance, this approach is used in~\cite{Chan1965,Robinson1961,Karimaki1991}.

However, since there is no analytic solution for minimisation of (\ref{eq:1}), one needs either to use iterative numerical methods or to change the loss function:
\begin{equation}
\label{eq:2}
\ln p(x_i, y_i | \theta)  = -\frac{1}{2}\ln 2 \pi \sigma - \frac{1}{2\sigma^2} \sum^{N}_{i=1} \left((x_i - x_0)^2 + (y_i - y_0)^2 - R^2\right)^2,
\end{equation}
as it is done in~\cite{Kasa1976,Delogne1972}.

We propose the following likelihood model:
\begin{equation}
\begin{aligned}
\label{eq:3}
\ln p(x_i, y_i, n_{x,i}, n_{y,i} | \theta) &= -\ln 2 \pi \sigma - \frac{1}{2\sigma^2} \sum^{N}_{i=1} \left[\left(n_{x,i}\left(x_0-x_i\right) + n_{y,i}\left(y_0-y_i\right) - R\right)^2 \right.\\ & + \left.\left(n_{y,i}\left(x_0-x_i\right) - n_{x,i}\left(y_0-y_i\right)\right)^2\right],
\end{aligned}
\end{equation}
where $n_{x,i},n_{y,i}$ is the measured normalised gradient at the point $x_i, y_i$.
The sign before $R$ is negative for outer disk edges and positive for inner annulus edges.
In essence, the gradient $n_{x,i},n_{y,i}$ is the normal vector at the circle point $x_i,y_i$.

Let us show how equation~(\ref{eq:3}) is derived.
We have the following four quantities for each point: $x_i,y_i,n_{x,i},n_{y,i}$.
Since the points are supposed to be located at the circle with some precision, the quantities are not independent random variables.
It is known that the vectors normal to the circle are crossed at its centre, so we assume that the following relations are held:
\begin{align}
n_{x,i} &= \frac{x_0 - x_i}{\sqrt{(x_0 - x_i)^2 + (y_0 - y_i)^2}} + \epsilon_{x,i}, \label{eq:nx_e}\\
n_{y,i} &= \frac{y_0 - y_i}{\sqrt{(x_0 - x_i)^2 + (y_0 - y_i)^2}} + \epsilon_{y,i}, \label{eq:ny_e}
\end{align}
where $\epsilon_{x,i}$ and $\epsilon_{y,i}$ are random variables with zero mean.
Consider the following two new random variables which geometrical meaning is explained in Fig.~\ref{fig:1}:
\begin{align}
d_{l} &= n_{x,i}\left(x_0-x_i\right) + n_{y,i}\left(y_0-y_i\right), \\
d_{\tau} &= n_{y,i}\left(x_0-x_i\right) - n_{x,i}\left(y_0-y_i\right).
\end{align}
Using equations~(\ref{eq:nx_e}) and~(\ref{eq:ny_e}), one may find mean values for $d_l$ and $d_{\tau}$:
\begin{align}
{\mathrm E}\left[d_l\right] &= {\mathrm E}\left[\sqrt{(x_0 - x_i)^2 + (y_0 - y_i)^2}\right]=R, \\
{\mathrm E}\left[d_{\tau}\right] &= 0,
\end{align}
where ${\mathrm E}\left[\cdot\right]$ denotes an expectation operator.
Here we assume that $\epsilon_{x,i}$ and $\epsilon_{y,i}$ are independent from $x_i$ and $y_i$.

There can also be an alternative geometric interpretation for~(\ref{eq:3}).
It is easy to show that every circle is located at some two-dimensional plane in the four-dimensional space of $x,y,n_x,n_y$.
Equation~(\ref{eq:3}) suggests minimal distances between the plane and the measurement points in this four-dimensional space.

The model has two important properties.
First, it can be reduced to the exact model~(\ref{eq:1}).
Indeed, expanding and collecting the terms in~(\ref{eq:3}) we have the following:
\begin{equation}
\begin{aligned}
\label{eq:3a}
\ln p(x_i, y_i, n_{x,i}, n_{y,i} | \theta) &= -\ln 2 \pi \sigma - \frac{1}{2\sigma^2} \sum^{N}_{i=1} \Bigl[ \left(x_0-x_i\right)^2 + \left(y_0-y_i\right)^2 + R^2 \Bigr.\\ & \Bigl.- 2 R n_{x,i} \left(x_0-x_i\right) - 2 R n_{y,i} \left(y_0-y_i\right) \Bigr],
\end{aligned}
\end{equation}
which is the same as~(\ref{eq:1}), when equations~(\ref{eq:nx_e}) and~(\ref{eq:ny_e}) are substituted in (\ref{eq:3a}), and $\epsilon_{x,i}=\epsilon_{y,i}=0$.
That corresponds to the case of unobservable normal vectors.

Second, it has the following simple analytic solution for the maximal likelihood estimators:
\begin{equation}
\label{eq:4a}
R = \frac{\frac{\sum^{N}_{i=1} n_{x,i}}{N} \frac{\sum^{N}_{i=1} x_i}{N} + \frac{\sum^{N}_{i=1} n_{y,i}}{N} \frac{\sum^{N}_{i=1} y_i}{N} - \frac{\sum^{N}_{i=1} x_i n_{x,i}}{N} - \frac{\sum^{N}_{i=1} y_i n_{y,i}}{N}}{1 - \left(\frac{\sum^{N}_{i=1} n_{x,i}}{N}\right)^2 - \left(\frac{\sum^{N}_{i=1} n_{y,i}}{N}\right)^2},
\end{equation}
\begin{equation}
x_0 = \frac{\sum^{N}_{i=1} x_i}{N} + R \frac{\sum^{N}_{i=1} n_{x,i}}{N},
\end{equation}
\begin{equation}
y_0 = \frac{\sum^{N}_{i=1} y_i}{N} + R \frac{\sum^{N}_{i=1} n_{y,i}}{N},
\end{equation}
\begin{equation}
\begin{aligned}
\label{eq:4d}
\sigma^2 &= \frac{1}{2N} \sum^{N}_{i=1} \Bigl[ \left(n_{x,i}\left(x_0-x_i\right) + n_{y,i}\left(y_0-y_i\right) - R\right)^2 \Bigr.\\
& \Bigl.+ \left(n_{y,i}\left(x_0-x_i\right) - n_{x,i}\left(y_0-y_i\right)\right)^2 \Bigr].
\end{aligned}
\end{equation}

In~\cite{Li2011}, a similar approach is used, authors implicitly constructed normal vectors from the model parameters and the measured point positions.
However, since they didn't measure the normal vectors, their disk parameters estimates are obtained through an iterative procedure.

\section{Maximal likelihood disk fitting}
\label{sec:eval}

\begin{figure}
\centering
\includegraphics[width=0.8\textwidth]{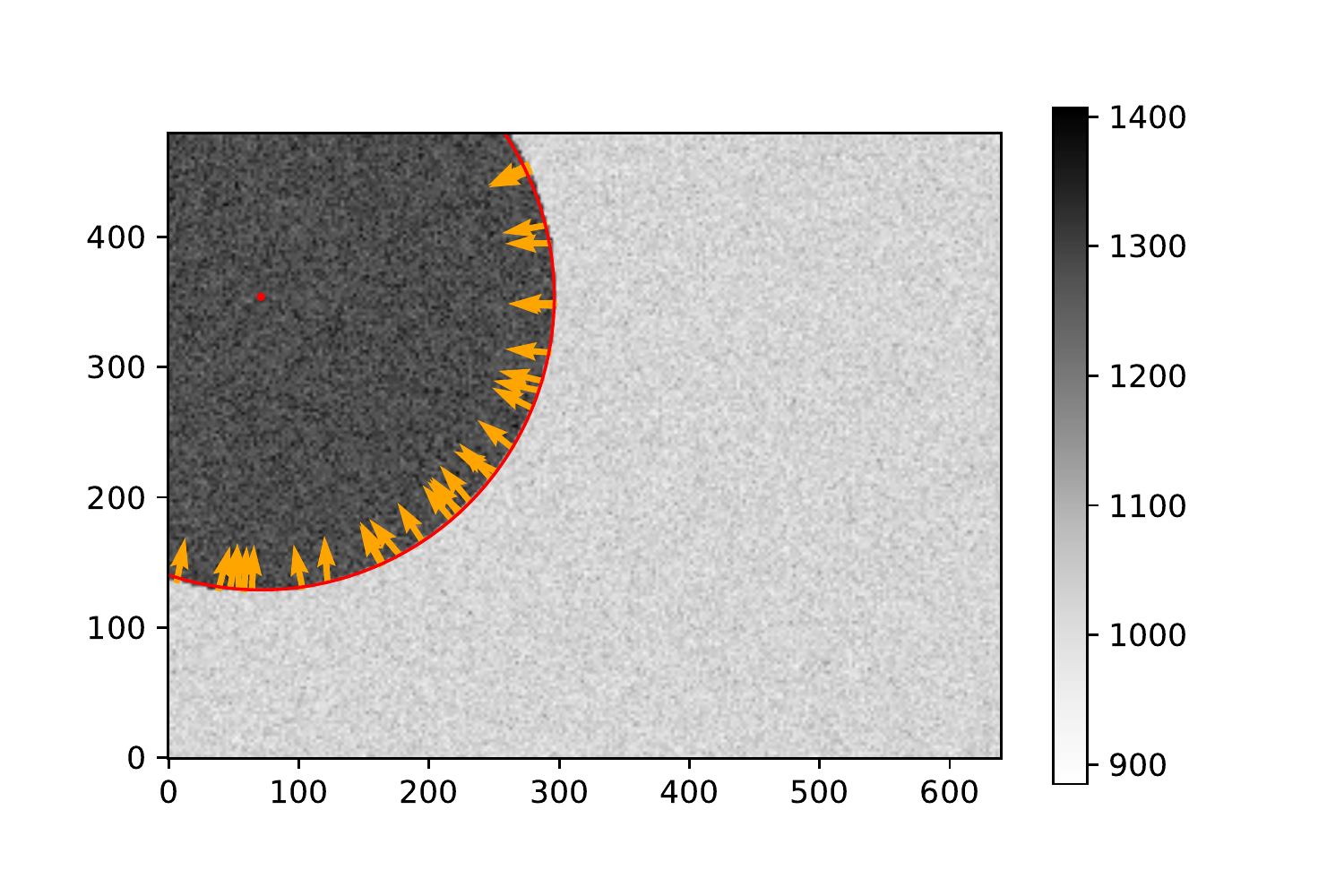}
\caption{\label{fig:0}
A sample simulated image.
The arrows denote gradient directions obtained by gradient filtering for random edge pixels.
The point and the arc represent disk parameters (centre and radius) fitted using described technique.
}
\end{figure}

Let us recall that we are interested in locating a single disk on a 2D image.
We suppose that the image is represented as a 2D grey scale array.
The following two steps are performed on the data to convert the array to the list of circle points coordinates suitable for applying equation~(\ref{eq:3}).

First, the gradient filters are applied to produce the horizontal gradient $g_{x,nm}$ and vertical gradient $g_{y,nm}$ component maps.
The following convolution kernels are used to evaluate the gradients:
\begin{align}
\label{eq:4}
K_{x,pq} &\equiv p \, \exp(-\frac{p^2}{2 s^2}) \exp(-\frac{q^2}{2 s^2}),\\
\label{eq:5}
K_{x,pq} &\equiv \exp(-\frac{p^2}{2 s^2}) \, q \, \exp(-\frac{q^2}{2 s^2}),\quad \left|p\right|,\left|q\right|\le 2.
\end{align}
where $s^2$ is selected to be $2$ to fit the kernel into five pixels.
We also ignore the Gaussian norm in the kernel, since we are interested in normalised gradient vectors rather than absolute gradient values.
Kernels~(\ref{eq:4}) and~(\ref{eq:5}) are separable\footnote{i.e. a product of two expressions each depending on its own index.}, so the filtering can be efficiently applied as a sequence of two 1D convolutions.
When the both gradient projections are known, the gradient norm $g_{nm} = \sqrt{g_{x,nm}^2+g_{y,nm}^2}$ map can be evaluated.

Second, Otsu threshold selection method~\cite{Otsu1979} is applied to the gradient norm map to detect pixels belonging to the circle (disk edge).
The thresholding technique relies on the data distribution itself, so the Gaussian norm in kernels~(\ref{eq:4})--(\ref{eq:5}) may be safely ignored here too.
All the image pixels are divided into the circle and the background using the threshold.
There may be too many first class pixels, so the smaller fixed-size subset is chosen at random.
Adjusting subset size helps to find a balance between computational expenses and accuracy.
Finally, for every pixel from the subset we know its integer coordinates $x_i=n, y_i=m$
and
normalised gradient vector~$n_{x,i}=g_{x,nm} \cdot g_{nm}^{-1}, n_{y,i}=g_{y,nm} \cdot g_{nm}^{-1}$.
Then, the gradient-based maximum likelihood algorithm is applied to the list to estimate circle parameters.

To evaluate the algorithm performance, we generate $10000$ sample disk images $640 \times 480$ pixels each.
Each disk has a random size and a random position.
The radius is distributed uniformly between $30$ and $270$ pixels.
The position is distributed uniformly between zero and the maximal dimension to allow existing of partial disk segments.
The disk pixels have a value of $255$, the background is $0$, and then
we apply Poisson noise of different scales $\lambda$ to the pixels of this input data set.
A sample is shown at Fig.~\ref{fig:0}.
For each combination of the noise level $\lambda$ and the points number each input image is processed as described above,
the circle parameters are estimated using equations~(\ref{eq:4a})--(\ref{eq:4d}), and an a-posteriori estimation error is evaluated.
The maximal absolute parameter difference is used as an accuracy measure.
The disk parameters are expressed in units of pixels.

The proposed algorithm is also compared to the algorithm of Li~et~al. (referred as Li's algorithm below) presented in~\cite{Li2011}.
Since that algorithm is iterative, the termination criterion is chosen to be so that relative error is less than $10^{-4}$ to ensure that the accuracy is limited by the data properties rather than an insufficient number of iterations.

\begin{table}[t!]
\begin{adjustwidth}{-1.5in}{-1.5in}
\begin{center}
\begin{tabular}{lr|lll|lll|lll|lll|lll}
\toprule
           & Points & \multicolumn{3}{c}{30} & \multicolumn{3}{c}{60} & \multicolumn{3}{c}{120} & \multicolumn{3}{c}{240} & \multicolumn{3}{c}{320} \\
           & Percentile &  25\% &  50\% &  75\% &  25\% &  50\% &  75\% &  25\% &  50\% &  75\% &  25\% &  50\% &  75\% &  25\% &  50\% &  75\% \\
Technique & Noise level &      &      &      &      &      &      &      &      &      &      &      &      &      &      &      \\
\midrule
\multirow{3}{*}{This paper} & 1    & 0.49 & 1.21 & 2.69 & 0.35 & 0.87 & 1.97 & 0.24 & 0.64 & 1.48 & 0.17 & 0.46 & 1.13 & 0.15 & 0.40 & 1.02 \\
           & 256  & 0.55 & 1.32 & 2.99 & 0.38 & 0.96 & 2.18 & 0.27 & 0.71 & 1.67 & 0.20 & 0.52 & 1.33 & 0.17 & 0.45 & 1.22 \\
           & 1024 & 0.75 & 1.87 & 4.25 & 0.51 & 1.35 & 3.23 & 0.37 & 1.00 & 2.61 & 0.28 & 0.76 & 2.29 & 0.24 & 0.68 & 2.21 \\
\cline{1-17}
\multirow{3}{*}{Li, et al.} & 1    & 0.11 & 0.25 & 0.46 & 0.08 & 0.18 & 0.35 & 0.06 & 0.13 & 0.27 & 0.05 & 0.10 & 0.22 & 0.04 & 0.09 & 0.21 \\
           & 256  & 0.11 & 0.24 & 0.46 & 0.08 & 0.18 & 0.34 & 0.06 & 0.13 & 0.26 & 0.04 & 0.10 & 0.21 & 0.04 & 0.09 & 0.21 \\
           & 1024 & 0.11 & 0.25 & 0.48 & 0.08 & 0.18 & 0.36 & 0.06 & 0.14 & 0.29 & 0.05 & 0.11 & 0.25 & 0.04 & 0.10 & 0.24 \\
\bottomrule
\end{tabular}

\caption{
\label{table:eval_center}
The algorithm evaluation results for the disk centre point $x_0,y_0$.
$25\%$, $50\%$, and $75\%$ percentiles are given for the absolute difference between the true and estimated parameters in units of pixels.
}
\end{center}
\end{adjustwidth}
\end{table}

\begin{table}[t!]
\begin{adjustwidth}{-1.5in}{-1.5in}
\begin{center}
\begin{tabular}{lr|lll|lll|lll|lll|lll}
\toprule
           & Points & \multicolumn{3}{c}{30} & \multicolumn{3}{c}{60} & \multicolumn{3}{c}{120} & \multicolumn{3}{c}{240} & \multicolumn{3}{c}{320} \\
           & Percentile &  25\% &  50\% &  75\% &  25\% &  50\% &  75\% &  25\% &  50\% &  75\% &  25\% &  50\% &  75\% &  25\% &  50\% &  75\% \\
Technique & Noise level &      &      &      &      &      &      &      &      &      &      &      &      &      &      &      \\
\midrule
\multirow{3}{*}{This paper} & 1    & 0.33 & 0.80 & 2.19 & 0.32 & 0.71 & 1.86 & 0.33 & 0.70 & 1.71 & 0.36 & 0.72 & 1.63 & 0.37 & 0.73 & 1.62 \\
           & 256  & 0.38 & 0.93 & 2.54 & 0.38 & 0.85 & 2.16 & 0.42 & 0.88 & 2.11 & 0.45 & 0.90 & 2.04 & 0.47 & 0.94 & 2.07 \\
           & 1024 & 0.62 & 1.50 & 4.00 & 0.64 & 1.45 & 3.76 & 0.72 & 1.50 & 3.54 & 0.80 & 1.59 & 3.70 & 0.81 & 1.63 & 3.65 \\
\cline{1-17}
\multirow{3}{*}{Li, et al.} & 1    & 0.09 & 0.21 & 0.42 & 0.07 & 0.15 & 0.31 & 0.05 & 0.11 & 0.25 & 0.04 & 0.08 & 0.21 & 0.03 & 0.07 & 0.20 \\
           & 256  & 0.09 & 0.20 & 0.41 & 0.07 & 0.14 & 0.30 & 0.05 & 0.11 & 0.23 & 0.04 & 0.08 & 0.20 & 0.03 & 0.07 & 0.21 \\
           & 1024 & 0.09 & 0.21 & 0.43 & 0.07 & 0.15 & 0.34 & 0.05 & 0.11 & 0.29 & 0.04 & 0.09 & 0.25 & 0.03 & 0.08 & 0.25 \\
\bottomrule
\end{tabular}

\caption{
\label{table:eval_radius}
The algorithm evaluation results for the disk radius $R$.
The table layout and units are the same as for Table~\ref{table:eval_center}.
}
\end{center}
\end{adjustwidth}
\end{table}

The evaluation results are given in Table~\ref{table:eval_center} and Table~\ref{table:eval_radius}.
We may see that the proposed algorithm has an important drawback.
It has a considerably lower accuracy than the algorithm from~\cite{Li2011} has.
The origin of this additional error is clear.
When we add normalised vectors~$(n_{x,i},n_{y,i})$ into consideration, we also introduce additional measurements error.
On the other hand, it takes only about $60 \times 10^{-6} s$ for the proposed algorithm to process one circle,
while it takes $2 \times 10^{-3} s$ for Li's algorithm to reach the obtained accuracy.
The execution times were measured at the Intel~i5-7200U~CPU for Python based implementation.
They are consistent within this paper but should be thought of as relative values.

The execution times can be made comparable by limiting the maximal iteration number for Li's algorithm.
The median error for Li's algorithm is about $20$ pixels in that case.
It is natural to try using the results of the proposed algorithm as initial estimates for iterative Li's algorithm.
In that case, Li's algorithm execution time is reduced to $0.9 \times 10^{-3} s$ keeping the same good accuracy as in Tables~\ref{table:eval_center} and~\ref{table:eval_radius}.

The evaluation results make us optimistic about the algorithm applications.
Since the proposed algorithm is not an iterative one, its major advance is the low and predictable execution time.

\section{Astronomical telescope pupil detection}
\label{sec:pupil}

Simplicity of equations (\ref{eq:4a})--(\ref{eq:4d}) providing optimal parameters for likelihood function~(\ref{eq:3}) allows us to 
construct more complex models on top of it.
The expectation maximisation~(EM) algorithm~\cite{Dempster1977} is usually considered as an elegant and effective way to maximise complex likelihood functions for so-called mixture models, when the probability density function for every observation is expressed as a linear combination of the single-class probability density functions each having its own parameters to be estimated.

When $N$ independent observations are given, the EM algorithm iteratively optimises the following goal function:
\begin{equation}
\label{eq:EM_Q}
Q\left(\bf{\theta}|{\theta'}\right) = \sum^{M}_{j=1} \sum^{N}_{i=1} T'_{i,j} \ln p_j({\mathbf x_i} | \theta),
\end{equation}
where $\theta$ is a distribution parameters vector to be estimated, ${\mathbf x_i}$ is a vector of $i$-th observation, index $j$ enumerates the class, and weights $T'_{i,j}$ are evaluated based on the previously estimated distribution parameters $\theta'$ as:
\begin{align}
T'_{i,j} &= \frac{p_j({\mathbf x_i} | \theta')}{\sum^{M}_{k=1} p_k({\mathbf x_i} | \theta')}.
\end{align}

The next step parameters estimate $\theta$ is evaluated as a solution for the following optimisation problem:
\begin{equation}
\theta = \arg \min_{\theta} Q\left(\bf{\theta}|{\theta'}\right).
\end{equation}
It is important to note that the equations for $\theta$ can be easily derived if a solution for every $j$-th single-class likelihood problem is known.
This property follows from the linearity of (\ref{eq:EM_Q}) and forms the key part of EM technique success.
If $p_j$ is described by~(\ref{eq:3}) for some $j$, then the corresponding parameters $x_{0}, y_{0}, R, \sigma$ are evaluated using modified equations~(\ref{eq:4a})--(\ref{eq:4d})
where $\sum^{N}_{i=1} T'_{i,j} \left(\cdot\right)_i$ is substituted for each $\sum^{N}_{i=1} \left(\cdot\right)_i$,
and $\sum^{N}_{i=1} T'_{i,j}$ is substituted for $N$.

\begin{figure}
\centering
\begin{tikzpicture}

\filldraw[pattern=north west lines,pattern color=gray,draw=black]
(0cm,-0.3cm) -- ++(0.125cm,0cm) to[bend right=10] ++(0cm,0.6cm) -- ++(-0.125cm,0cm) -- cycle;

\filldraw[pattern=north west lines,pattern color=gray,draw=black]
(4cm,-1cm) -- ++(0.25cm,0cm) -- ++(0, 0.75cm) -- ++(-0.125cm,0cm) to[bend left=5] cycle;

\filldraw[pattern=north west lines,pattern color=gray,draw=black]
(4cm,1cm) -- ++(0.25cm,0cm)  -- ++(0, -0.75cm) -- ++(-0.125cm,0cm) to[bend right=5] cycle;

\draw[very thin, postaction={on each segment={mid arrow=black!75}}, color=black!75]
(-1cm,-1cm) -- (4cm,-1cm)
--
(0.125cm,-0.3cm) -- (6cm,0.0625cm);

\draw[very thin, postaction={on each segment={mid arrow=black!75}}, color=black!75]
(-1cm,1cm) -- (4cm,1cm)
--
(0.125cm,0.3cm) -- (6cm,-0.0625cm);

\end{tikzpicture}
\caption{
\label{fig:2}
Ritchey-Chretien optical telescope scheme.
The right-most is the primary mirror, the left-most is the secondary mirror, both having hyperbolic surface.
The primary mirror has a central hole for the passing light beams to the focal plane.
}
\end{figure}
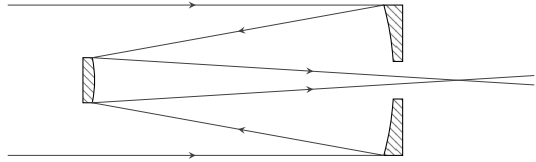

\begin{figure}
\centering
\begin{tabular}{cc}
\includegraphics[width=0.4\textwidth]{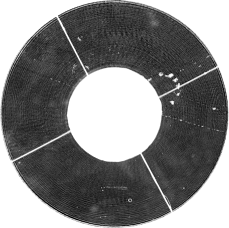} &
\includegraphics[width=0.4\textwidth]{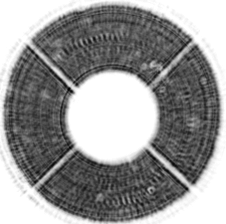}
\end{tabular}
\caption{\label{fig:3}
Left panel: an inverted image of the entrance pupil of a telescope.
Right panel: the same when defocused.
Central circular white spot is the shadow from the secondary mirror.
Lines going through the image are the shadows from the spiders, constructions carrying the secondary mirror.
The images have been obtained at 2.5~m telescope of Caucasian Mountain Observatory, Lomonosov Moscow State University.
}
\end{figure}

Let us consider the following real-data case as an example for applying the EM
algorithm on top of the proposed disk parameters estimation algorithm.

An astronomical telescope focuses parallel beams of the light coming from stars at a focal plane.
Usually, two-mirror optical schemes are common nowadays.
A typical optical scheme for a modern optical astronomical telescope is given at Fig.~\ref{fig:2}.
Note, that the part of the parallel light beam is dissipated by the secondary mirror back surface.
The outer size of the beam is limited by the primary mirror size.
An optical scheme part limiting the amount of light passing through the whole system is called an actual aperture stop.
We may imagine that the aperture stop splits the whole optical system into the two parts.
An image of the actual aperture stop produced by the back of the optical system is called an exit pupil.
An entrance pupil is the image of the actual aperture stop produced by the front optical system part in the reverse direction.
The entrance pupil is identical to the actual aperture stop in case of typical astronomical telescope optical scheme given at Fig.~\ref{fig:2}.
The pupil image can be obtained by putting an image detector at the exit pupil.
The sample for the real telescope is given at Fig.~\ref{fig:3}.

Installing additional equipment in the exit pupil is used, for instance, while examining the telescope optics quality~\cite{Potanin2009,Potanin2017} or optical turbulence profiling~\cite{Fuchs1998}.
Since the telescope optical system geometry is always a subject of thermal expansions and gravitational bending, the exact position for the exit pupil is known with limited accuracy.
In real conditions, an additional adjustment would be required.
The sharpness of the outer pupil edge may be controlled to achieve the goal.
Note, that the inner pupil edge can be unsharp at the same moment since it is produced by the secondary mirror edge which is located at some distance from the primary mirror.
This is why the outer edge locating, followed by spatial filtering, is required before sharpness detection.

The proposed algorithm naturally distinguishes between an annulus outer edge and inner one.
The outer edge has gradients directed towards the circle centre where they are crossed.
At the same time, the inner edge has gradients directed in the reverse direction.
We could craft an algorithm detecting inner edges by changing the sign for $R$ in equation~(\ref{eq:3}), assuming that $R$ is always positive.

Two classes of edge points are obtained by the edge detection algorithm in the case of images similar to Fig.~\ref{fig:3}.
The first class is consisted from the outer edge points.
The second class consists of the inner edge points and the spiders edge points.
We assume that the gradient vectors direction is distributed uniformly for the latter class.
Since the class is a hidden variable, it seems to be natural to apply mentioned expectation maximisation algorithm~\cite{Dempster1977} on the following two-component model.
The probability density functions are defined as follows:
\begin{equation}
\label{eq:20}
\begin{aligned}
p_1(x_i, y_i, n_{x,i}, n_{y,i} | {\theta}) &= \frac{\tau}{2\pi\sigma^2_1}\exp\biggl(-\frac{1}{2\sigma_1^2}\left(\left(n_{x,i}\left(x_{0,1}-x_i\right) + n_{y,i}\left(y_{0,1}-y_i\right) - R\right)^2\right.\biggr. \\ &+ \biggl.\left.\left(n_{y,i}\left(x_{0,1}-x_i\right) - n_{x,i}\left(y_{0,1}-y_i\right)\right)^2\right)\biggr), \\
\end{aligned}
\end{equation}
\begin{equation}
\label{eq:21}
p_2(x_i, y_i | {\theta}) = \frac{1-\tau}{2\pi\sigma^2_2}\exp\left(-\frac{1}{2\sigma_2^2}\left(\left(x_{0,2}-x_i\right)^2 + \left(y_{0,2}-y_i\right)^2 \right)\right).
\end{equation}
Here, $p_1$ follows from equation~(\ref{eq:3}) and represents the outer edge point distribution.
The vector ${\theta}$ stands for the probability density function parameters to be determined: $\tau, x_{0,1}, y_{0,1}, R, \sigma_1, x_{0,2}, y_{0,2}, \sigma_2$.
The parameters are evaluated using modified equations~(\ref{eq:4a})--(\ref{eq:4d}) and ordinarily equations for the Gaussian mixture EM model.

The initial parameter estimation is required to start the iteration process.
The following has been found to be working satisfactory.
The pupil centre is estimated as the solution for the following linear equation set:
\begin{align}
x_0 \sum^{N}_{i=1} {n_{y,i}^2} - y_0 \sum^{N}_{i=1} {n_{x,i} n_{y,i}} &= \sum^{N}_{i=1} n_{y,i} \left(n_{y,i} x_i - n_{x,i} y_i\right), \\
x_0 \sum^{N}_{i=1} {n_{x,i} n_{y,i}} - y_0 \sum^{N}_{i=1} {n_{x,i}^2} &= \sum^{N}_{i=1} n_{x,i} \left(n_{y,i} x_i - n_{x,i} y_i\right),
\end{align}
while other initial parameters are evaluated as follows:
\begin{align}
x_{0,1} = x_{0,2} &= x_0, \\
y_{0,1} = y_{0,2} &= y_0, \\
\sigma_2 = R &= \sqrt{\frac{\sum^{N}_{i=1}\left((x_i - x_{0})^2 + (y_i - y_{0})^2\right)}{N}}, \\
\sigma_1 &= \text{some sufficiently large constant}.
\end{align}

We examined two data sets of real data: one for the focused pupil image and another for the defocused pupil.
Each data set consists of $1000$ individual images obtained in a series under the same conditions.
The pupil parameters estimator accuracy can be evaluated by running the algorithm for every individual image frame followed by evaluating the parameters standard deviation.

In the focused pupil case, we have found that the standard deviation for the pupil centre $\sigma_c = 0.85$~pix. and the standard deviation for the outer pupil radius $\sigma_R = 0.42$~pix.
Given the average outer radius $R = 113.5$~pix, we have a relative accuracy of $0.4\%$ here.
For the defocused pupil case, $\sigma_c = 1.2$~pix. and $\sigma_R = 0.82$~pix, while the average outer radius remains almost the same $R = 113.4$~pix.

It takes approximately $120 \times 10^{-3} s$ to process a single $659 \times 493$ frame using Python programming language in both cases.
Most of the time is spent at the gradient filtering stage.
However, our production C++ based algorithm implementation requires only $6 \times 10^{-3} s$ that allows us to process approximately $160$~frames per second.

Hence, we have found that the proposed algorithm is able to cope with real data.

\section{Conclusion}
In this paper, the new likelihood disk fitting model was presented.
We showed how using additional information about image gradients can change the model properties.
We obtained likelihood model~(\ref{eq:3}) that
has simple equations~(\ref{eq:4a})--(\ref{eq:4d}) for its solution
and
can be reduced to the exact likelihood model~(\ref{eq:1}) at the same time.
The likelihood model parameters evaluating doesn't require iterative techniques because of the equations' simplicity.
So, compared to iterative techniques, less computational resources are generally required,
and the computational time is more predictable, which may be important in real-time image processing.

At the same time, the proposed model also has the following considerable drawbacks.
First, additional input data are required.
However, when gradient-based edge detection algorithms are applied during the image processing pipeline, the required data is usually already available.
Second, it can only apply to disks or annuluses.
Fitting real circles may require further algorithm complication, such as considering a circle as a thin annulus and using expectation maximisation technique to fit the annulus parameters using both its outer and inner edges.
Third, it has a lower accuracy than can be reached by other algorithms under the same conditions.

In Sections~\ref{sec:eval} and~\ref{sec:pupil}, we demonstrated that the proposed likelihood model may be efficiently used both solely and for estimating initial parameters for iterative circle fitting algorithms.

\bibliography{biblio} 

\begin{thebibliography}{10}

\bibitem{Kasa1976}
I.~{Kasa}, ``A circle fitting procedure and its error analysis,'' {\em IEEE
  Transactions on Instrumentation and Measurement}, vol.~IM-25, pp.~8--14,
  March 1976.

\bibitem{Karimaki1991}
V.~Karimäki, ``Effective circle fitting for particle trajectories,'' {\em
  Nuclear Instruments and Methods in Physics Research Section A: Accelerators,
  Spectrometers, Detectors and Associated Equipment}, vol.~305, no.~1, pp.~187
  -- 191, 1991.

\bibitem{Crawford1983}
J.~{Crawford}, ``A non-iterative method for fitting circular arcs to measured
  points,'' {\em Nuclear Instruments and Methods in Physics Research},
  vol.~211, no.~1, pp.~223 -- 225, 1983.

\bibitem{Lin2004}
D.-T. {Lin} and C.-M. {Yang}, ``Real-time eye detection using face-circle
  fitting and dark-pixel filtering,'' in {\em 2004 IEEE International
  Conference on Multimedia and Expo (ICME)}, vol.~2, pp.~1167--1170, June 2004.

\bibitem{Landau1987}
U.~Landau, ``Estimation of a circular arc center and its radius,'' {\em
  Computer Vision, Graphics, and Image Processing}, vol.~38, no.~3, pp.~317 --
  326, 1987.

\bibitem{Coach2003}
G.~{Coath} and P.~{Musumeci}, ``{Adaptive arc fitting for ball detection in
  RoboCup},'' in {\em {APRS Workshop on Digital Image Computing}}, (Brisbane,
  QLD, Australia), pp.~63 -- 68, Feb. 2003.

\bibitem{Nunez2008}
P.~{Nunez}, R.~{Vazquez-Martin}, A.~{Bandera}, and F.~{Sandoval}, ``An
  algorithm for fitting 2-d data on the circle: Applications to mobile
  robotics,'' {\em IEEE Signal Processing Letters}, vol.~15, pp.~127--130,
  2008.

\bibitem{Zhang2006}
S.~{Zhang}, L.~{Xie}, and M.~D. {Adams}, ``Feature extraction for outdoor
  mobile robot navigation based on a modified gauss–newton optimization
  approach,'' {\em Robotics and Autonomous Systems}, vol.~54, no.~4, pp.~277 --
  287, 2006.

\bibitem{Ballard1981}
D.~{Ballard}, ``Generalizing the hough transform to detect arbitrary shapes,''
  {\em Pattern Recognition}, vol.~13, no.~2, pp.~111 -- 122, 1981.

\bibitem{Illingworth1987}
J.~{Illingworth} and J.~{Kittler}, ``The adaptive hough transform,'' {\em IEEE
  Transactions on Pattern Analysis and Machine Intelligence}, vol.~PAMI-9,
  pp.~690--698, Sep. 1987.

\bibitem{Atherton1999}
T.~Atherton and D.~Kerbyson, ``Size invariant circle detection,'' {\em Image
  and Vision Computing}, vol.~17, no.~11, pp.~795 -- 803, 1999.

\bibitem{Zelniker2006}
E.~E. {Zelniker} and I.~V.~L. {Clarkson}, ``Maximum-likelihood estimation of
  circle parameters via convolution,'' {\em IEEE Transactions on Image
  Processing}, vol.~15, pp.~865--876, April 2006.

\bibitem{Otsu1979}
N.~{Otsu}, ``A threshold selection method from gray-level histograms,'' {\em
  IEEE Transactions on Systems, Man, and Cybernetics}, vol.~9, pp.~62--66, Jan
  1979.

\bibitem{Chan1965}
N.~N. Chan, ``On circular functional relationships,'' {\em Journal of the Royal
  Statistical Society: Series B (Methodological)}, vol.~27, no.~1, pp.~45--56,
  1965.

\bibitem{Robinson1961}
S.~M. Robinson, ``Fitting spheres by the method of least squares,'' {\em
  Commun. ACM}, vol.~4, pp.~491--, Nov. 1961.

\bibitem{Delogne1972}
P.~Delogne, ``Computer optimization of deschamps’ method and error
  cancellation in reflectometry,'' in {\em Proc. IMEKO-Symp. Microwave
  Measurements}, pp.~117--123, 1972.

\bibitem{Li2011}
W.~{Li}, J.~{Zhong}, T.~A. {Gulliver}, B.~{Rong}, R.~Q. {Hu}, and Y.~{Qian},
  ``Fitting noisy data to a circle: A simple iterative maximum likelihood
  approach,'' in {\em 2011 IEEE International Conference on Communications
  (ICC)}, pp.~1--5, June 2011.

\bibitem{Dempster1977}
A.~P. {Dempster}, N.~M. {Laird}, and D.~B. {Rubin}, ``Maximum likelihood from
  incomplete data via the em algorithm,'' {\em Journal of the Royal Statistical
  Society: Series B (Methodological)}, vol.~39, no.~1, pp.~1--22, 1977.

\bibitem{Potanin2009}
S.~A. {Potanin}, ``Shack-hartmann wavefront sensor for testing the quality of
  the optics of the 2.5-m sai telescope,'' {\em Astronomy Reports}, vol.~53,
  pp.~703--709, Aug 2009.

\bibitem{Potanin2017}
S.~A. {Potanin}, I.~A. {Gorbunov}, A.~V. {Dodin}, A.~D. {Savvin}, B.~S.
  {Safonov}, and N.~I. {Shatsky}, ``Analysis of the optics of the 2.5-m
  telescope of the sternberg astronomical institute,'' {\em Astronomy Reports},
  vol.~61, pp.~715--725, Aug 2017.

\bibitem{Fuchs1998}
A.~{Fuchs}, M.~{Tallon}, and J.~{Vernin}, ``{Focusing on a Turbulent Layer:
  Principle of the ``Generalized SCIDAR''},'' {\em Publications of the
  Astronomical Society of the Pacific}, vol.~110, pp.~86--91, Jan. 1998.

\end{thebibliography}
\bibliographystyle{ieeetr}

\end{document}